\documentstyle[preprint,aps]{revtex}
\input{psfig}
\newcommand{\be}{\begin{equation}}
\newcommand{\ee}{\end{equation}}
\newcommand{\ba}{\begin{eqnarray}}
\newcommand{\ea}{\end{eqnarray}}
\begin{document}
\title{Sound waves and
the absence of Galilean invariance in flocks}
\author{Yuhai Tu}
\address{IBM T. J. Watson Research Center, P. O. Box 218, Yorktown Heights, NY 10598}
\author{John Toner and Markus Ulm}
\address{Department of Physics, University of Oregon,
Eugene, OR 97403-1274}

\maketitle

\begin{abstract}
We study a model of flocking for a very large system (N=320,000)
numerically. We find that in the long wavelength, long time
limit, the fluctuations of the velocity and density fields are carried by propagating sound modes, whose dispersion and damping agree quantitatively with the predictions of
our previous work using a continuum
equation. 
We find that the sound velocity is anisotropic and characterized by its speed $c$ for propagation perpendicular to the mean velocity $<\vec{v}>$, $<\vec{v}>$ 
itself, and a third velocity $\lambda <\vec{v}>$, arising explicitly from the lack of Galilean invariance in flocks.
\end{abstract}

\newpage

The dynamics of ``flocking" behavior
of living things, such as birds, fish, wildebeest, slime
molds and bacteria has long attracted a great deal of attention among
biologists, computer animators and physicists\cite{boids,patridge,Vicsek}. 
It is crucial to correctly describe the interaction
between members of the flock
in order to understand and model the flocking behavior. 
As summarized in [2], a large flock
does not have a global leader; instead, the impressive collective flocking phenomena is
caused by individual members of the flock following the motion of
their neighbors. 

In our earlier work\cite{us}, we studied the flocking
dynamics by using continuum equations for the coarse-grained density field
$\rho(\vec{x},t)$ and velocity field $\vec{v} (\vec{x},t)$, written as: 

\ba
\partial_{t} \vec{v}+\lambda (\vec{v}\cdot\nabla)\vec{v}= &\alpha& \vec{v}-\beta
|\vec{v}|^{2}\vec{v}
-\nabla P
+D_L \nabla (\nabla
\cdot \vec{v})\nonumber\\
&+& D_{1}\nabla^{2}\vec{v}+
D_{2}(\vec{v}\cdot\nabla)^{2}\vec{v}+\vec{f}\\
\frac{\partial\rho}{\partial t}+\nabla\cdot(\vec{v}\rho)=&0&
\ea
where $\beta$, $D_{1}$, $D_{2}$ and $D_{L}$ are all positive, and $\alpha < 0$
in the disordered phase and $\alpha>0$ in the ordered state. 
The $\alpha$ and $\beta$ terms simply make the local $\vec{v}$ have a non-zero
magnitude $(=\sqrt{\alpha/\beta})$ in the ordered phase. $D_{L,1,2}$
are diffusion constants.
The Gaussian random noise
$\vec{f}$ has
correlations:
$
<f_{i}(\vec{r},t)f_{j}(\vec{r}',t')>=\Delta \delta_{ij}\delta^{d}(\vec{r}-\vec{r}')\delta(t-t')
$
where $\Delta$ is a constant, and $i$ , $j$ denote Cartesian components.
Finally, the pressure
$P=P(\rho)=\sum_{n=1}^{\infty} \sigma_n (\rho-\rho_0)^n , $
where $\rho_0$ is the mean of
the local number density $\rho(\vec{r})$ and $ \sigma_n$ are coefficients
in the pressure expansion.
The final equation (2) reflects conservation of birds.

In [4], we considered the special case of (1) with $\lambda=1$.
Just as the absence of the Galilean invariance for
the flock motion allows $\alpha$ and $\beta\ne 0$ in equation (1), 
likewise $\lambda$ need not be $=1$. In [5] and this paper, we consider the more generic case $\lambda\ne 1$, which leads to a different direction dependence of
the sound speed than when $\lambda=1$ \cite{foot1}.

In the ordered phase where $\alpha>0$, the velocity field and the density field can be 
written as: $\vec{v}=v_{s}\hat{x}_{||}+\vec{\delta v}$, $\rho=\rho_0+\delta \rho$
where $\rho_0$ and $v_{s}\hat{x}_{||}$ are the space averaged density and 
spontaneous symmetry broken velocity respectively. 
The spontaneous symmetry
breaking of a vector field leads to large ``Goldstone mode" fluctuations;
in flocks, this mode
is $\vec{v_{\perp}}$, the projection of $\delta \vec{v}$ perpendicular to 
$\hat{x}_{||}$ (we will hereafter use $``||"$ ($`` \perp"$) to denote the projection
of any vector along (perpendicular to) $\hat{x_{||}}$). Indeed, for 
equilibrium systems,
such fluctuations are 
strong enough in 
two dimensions to destroy the long range order \cite{mw}. 
One of the remarkable predictions of
our continuum model of flocking 
is that the ordered state is
stable even in two dimensions due to the non-equilibrium effect of the 
nonlinear terms. The mean squared fluctuations in Fourier space in 2D are: 

\ba
< | \delta \rho ({\vec q}, \omega) |^2 >&=&{\Delta q_\perp^2  \rho_0^2
\over S(\vec{q}, \omega)}\\
< | v_{\perp} ({\vec q}, \omega) |^2 >&=&{\Delta [(\omega-v_s q_{||})^2
+D_{\rho}q_{||}^4]
\over S(\vec{q}, \omega)}
\ea
where the denominator $S(\vec{q},\omega)=
[(\omega-v_s q_{||})(\omega-\lambda v_s q_{||})-c^2 q_\perp^2]^2+
[(\omega-v_s q_{||})(D^{R}_{\perp} (\vec{q})
q_{\perp}^2 + D_{||} q_{||}^2)-\lambda v_{s}D_{\rho}q_{||}^3]^2$.
$c=\sqrt{\sigma_1 \rho_0}$, $D_{\rho}=c^2/\alpha$, 
$D_{||}=D_1+D_2+D_{\rho}$ 
and $D^R_\perp$, the
renormalized diffusion constant, scales as:
\be
D^{R}_\perp (
{\vec q_\perp}, q_{||}; \lambda, \rho_0, \sigma_n)
=q_\perp^{z-2}
f ( {q_{||}\over q_\perp^\zeta})\;\;\;,
\ee
where the exponents $z$ and $\zeta$ are found by RG analysis to be 
$z=\frac{6}{5}$
and $\zeta=\frac{3}{5}$ for two dimensions, and the
scaling function $f(x)$ is universal up to an overall, 
nonuniversal $\vec{q}$ and $\omega$ independent scale factor.

From the above expressions, the correlation functions will have peaks around
$\omega_0(\vec{q})$ which satisfies:
$
(\omega_0-v_s q_{||})(\omega_0-\lambda v_s q_{||})-c^2 q_\perp^2=0
$ 
with solutions: $\omega_0=\Omega_{\pm}(\vec{q})=\frac{1}{2}(1+\lambda)v_s q_{||}
\pm (\frac{1}{4}(1-\lambda)^2 v_s^2 q_{||}^2+c^2q_{\perp}^2)^{1/2}$. 
This implies that for the wavevector $(q_{||},q_{\perp})=q(\cos
(\theta_q), \sin(\theta_q))$ at small $q$ , there should be two peaks 
in the power spectrum located around 
$ \omega_0 =c_{\pm}(\theta_q)q$ with 
the sound speeds:
\ba
c_{\pm}(\theta_q)&=&\frac{1}{2}(1+\lambda)v_s \cos(\theta_q)\nonumber\\ 
&\pm& (\frac{1}{4}(1-\lambda)^2 v_s^2 \cos^2 (\theta_q)+c^2\sin^2 
(\theta_q))^{1/2}
\ea
The relative strength of the two peaks varies with $\theta_q$.
It is not hard to see from eqs. (3), (4), that at $\theta_q\sim 0$ 
, $< | \delta \rho ({\vec q}, \omega) |^2 >$ will
only have a peak with corresponding wave velocity $c_{+}(\theta_q =0)=v_s$, 
and $< | v_{\perp} ({\vec q}, \omega) |^2 >$ will only have a peak at $
c_{-}(\theta_q=0)=\lambda v_s$.

In this paper, we study a discrete
model numerically to test the predictions made by our continuum theory. 
The model we use is very similar to 
the one studied
by Vicsek et al \cite{Vicsek}. Following [1], we call our simulated flockers
``boids". At a given time t, the position and 
the direction of the velocity
for each boid are given as $(\vec{r}_i(t)$,$ \theta _i(t)$)
for $i=1,2,\cdots,N$. The magnitude of the velocity is fixed:
$|\vec{v}_i|=v_0$, its direction is updated at the next time step 
by averaging over its neighbors' moving directions:
\be
\theta_i(t+1)=\Theta(\frac{1}{M}\sum_{j=1}^{M}(\vec{v}_j(t)+\vec{g}_{ij}(t))
+\vec{\eta_i}(t))
\ee
$M$ is the number of neighbors for boid $i$ within radius $R$: 
$r_{ij}=|\vec{r_i}-\vec{r_j}|<R$. The extra interaction term $\vec{g}_{ij}=
g_0 (\vec{r_i}-\vec{r_j})((\frac{l_0}{r_{ij}})^3-(\frac{l_0}{r_{ij}})^2))$
makes boids repel each other when they are closer than $l_0$, and
attract each other otherwise, with $l_0$ the average distance
between boids in the flock, 
this interaction will prevent formation of 
clusters. 
The noise term $\vec{\eta_i}(t)=\Delta v
(\cos (\pi e_i(t)), \sin (\pi e_i(t)))$,
where $e_i(t)$ is a random number in the interval $[-1,1]$. The function
$\Theta (\vec{x})$ is just the polar angle of the vector $\vec{x}$.
The position update
is simply: $\vec{r_{i}}(t+1)=\vec{r_i}(t)+v_0 (\cos(\theta _i(t)), \sin(\theta_
i(t)))$.
The parameters in this model are $R$, $\gamma$, $\Delta v$, $v_0$ and $g_0$.

The particular form of the interactions should not affect the universal predictions of the continuum theory presented above, but rather should only change
non-universal phenomenological parameters like $c$, $\lambda$, $D_{||}$ etc. .
They also affect the length scale $l_{NL}$ beyond which the asymptotic long wavelength forms of the correlation
functions (3) and (4) apply. 
Indeed, an one-loop RG analysis predicts: 
$l_{NL}\sim (10D_{\perp}^{\frac{5}{4}}D_{||}^{\frac{1}{4}}/\lambda 
\Delta ^{\frac{1}{2}})^{\frac{2}{4-d}}\times O(1)$. Higher loop corrections may
affect this result, but it presumably remains accurate to factors of $O(1)$.

For our numerical model, we estimate (on dimensional grounds): $\lambda\sim 1$,
$\Delta\sim (\Delta v)^2\frac{R^d}{t_0}$, $D_{||}\sim D_{\perp}\sim 
\frac{R^2}{t_0}$. Inserting these estimates, we find 
$l_{NL}\sim R( \frac{10R}{\Delta v t_{0}})^{\frac{2}{4-d}}$. In our simulation, choosing units of length and time such
that $R=t_{0}=1$, and taking $\Delta v \sim \Delta v_{c}\sim 1/3$ in 
these units, for $d=2$ we get the lower bound $\l_{NL}> 30$. Previous 
simulations \cite{vic2} took $\Delta v<<1$, and therefore have a much larger $l_{NL}$. 
Hence, no non-trivial nonlinear
effects could be observed since their systems were much smaller than $l_{NL}$.

The above analysis shows that in order to test the scaling 
behavior with a reasonable system size,
one seeks a small $l_{NL}$ by increasing $\Delta v$ and decreasing
the radius of interaction $R$ as much as possible 
without entering the disordered phase.
In this paper, 
we report the results of a simulation with system size $L\times L$ with $L=400$ 
and
the number of boids $N=320,000$. 
We choose $R=1$, $g_0=0.6$, $v_0=1.0$, $\l_0=.707$. For these parameter values, the
flock becomes disordered at $\Delta v_{c}\sim .375$ as shown in fig. 1(a). 
The order parameter $\phi$ is defined 
simply as the magnitude of the average velocity of the 
whole flock:$\phi=\frac{1}{N}|
\sum_{i=1}^{N}\vec{v_i}|$. 
To stay in the ordered 
phase and have enough fluctuations, 
we choose $\Delta v=0.15$. 

Previous simulations have used periodic boundary conditions \cite{Vicsek}. 
However, for any finite flock, the direction of
the average velocity will slowly change, making comparison to the analytical
results, which assume infinite system size and hence a constant 
direction for $<\vec{v}>$, difficult. In order to make $<\vec{v}>$ 
constant in its direction, we impose
periodic boundary conditions in one of the directions, say the $x$ direction, 
and reflecting boundary conditions in the other direction $y$, i.e., when
a boid $i$ with velocity $(v_i^{x},v_i^{y})$ collides with the "walls" at 
$y=\pm L/2$, its velocity changes to $(v_i^{x},-v_i^{y})$. 
The symmetry broken velocity is thus forced to lie along the $x$-direction, 
without
changing the bulk dynamics of the system. We will hereafter use ``$||$" and x; ``$\perp$"
and y interchangeably.


We first measure the equal time correlation functions. From eq. (3), we 
predict:
\ba 
C_{\rho}(\vec{q})&=&< \delta \rho ({\vec q}, t) \delta \rho (- \vec{q},t)  >
=\int < | \delta \rho ({\vec q}, 
\omega) |^2 > \frac{d\omega}{2\pi}\nonumber\\&=&\frac{2\Delta\rho_{0}^{2}}{c^2(D^{R}_{\perp} (\vec{q})
q_\perp^2 + D_{||} q_{||}^2)}
\ea
We see that the equal time correlation function gives 
us a direct measure of the attenuation. The asymptotic behavior of $C_{\rho}(\vec{q})$ can be 
expressed as:
\ba
C_{\rho}(\vec{q})&\sim q_{\perp}^{-z}, \;\;\;\;  q_{\perp}>>q_{||}^{\frac{1}{\zeta}}\\
&\sim q_{||}^{-z/ \zeta}, \;\;\;\; q_{\perp}<<q_ {||}^{\frac{1}{\zeta}}
\ea

In fig. 1(b),
we have plotted the equal time density correlation functions in Fourier
space: $ C_{\rho} (q_{||},q_{\perp}=2\pi /L)$  versus $q_{||}$
and $C_{\rho} (q_{||}=0, q_{\perp})$ versus $q_{\perp}$ from our simulation. 
The scaling behavior at long length scales can be fitted with: $
C_{\rho} (q_{||},q_{\perp}=2\pi /L)\sim q_{||}^{-2.05}$ and
$C_{\rho} (q_{||}=0, q_{\perp})\sim q_{\perp}^{-1.23}$.
These two exponents show excellent agreement with the analytical results
$-2$ and $-\frac{6}{5}$ respectively. 
As can be seen from fig. 1(b), the scaling region for the current 
simulation covers slightly less than one decade in $q_{\perp}$. It is not 
surprising that earlier simulations of
smaller systems with less carefully chosen parameters (leading to larger $l_{NL}$), did not observe the nontrivial scaling.

Another interesting measurement of the simulation is the anomalous diffusion
of individual boids in the direction $y$ perpendicular to the flock's 
moving direction.
We measure the ``width" of the dispersion of an ensemble of boids:
$ w^2(t)=<(y_i(t)-y_i(0))^2>$. 
The analytical behavior of the anomalous diffusion can be
obtained from:
$w^2(t)\sim \int_{0}^{t}\int_{0}^{t}<v_{y}^i (t')v_{y}^i(t'')>dt'dt''$
where $v_y^i(t)$ is the velocity of the $i$th boid along y direction at time t. The 
velocity correlation function is given by (4):
\ba
&<&v_{y}^i(0)v_{y}^i(t)>\sim<v_y(\vec{x}+\phi\hat{x}t,t)v_y(\vec{x},0)>\nonumber\\&=&\int {\exp(i(\omega-\phi q_{||}) t) 
\Delta (\omega-v_s q_{||})^2 d^2q d\omega
\over S(\vec{q},\omega) } 
\sim t^{1-1/\zeta}
\ea
which implies:
$w^2(t)\sim t^{3-1/\zeta}=t^{\frac{4}{3}}$. In fig. 2(b), we have plotted the
width squared $w^{2}(t)$ versus time $t$ in log-log scale. The scaling can
be fitted nicely with $w^2(t)\sim t^{1.3}$, which agrees well with the 
analytical result $t^{\frac{4}{3}}$. 
We have also simulated Vicsek's original model, but with
parameters $\Delta v$ etc. chosen to make $l_{NL}$ as small as possible, 
and found again $w^{2}(t)\sim t^{1.3}$. This
supports the universality of our analytic results.

Besides the scaling behavior, the analytical results (3), (4) also imply 
the existence of sound waves as reflected in
the peaks of
the correlation functions eqs. (3), (4). From eq. (3), at a given
value of $\vec{q}$, the correlation function has peaks at $\omega=c_{\pm}(\theta_q) q$.  
We have measured the power spectrum in the y-direction:  
$<|\delta\rho (q_{||}=0,q_{\perp}=\frac{2\pi}{L}n_{\perp} ,
\omega=\frac{2\pi}{T}n_{\omega})|^2>$ (T=1024) 
with different values of $n_{\perp}(=1,2,\cdots,20)$. 
Figure 3 shows the 
power spectra for $n_{\perp}=5,10,20$. The spectra are 
symmetric around $\omega=0$ (we only show half of the spectrum for $\omega>0$) 
and the positions of the peaks $n_{\omega}^{*}$ versus $n_y$ are shown in the
inset of figure 3, whose slope determines the sound velocity in the 
y-direction $c=0.62$. 
We have calculated the power spectrum of $v_{\perp}$ in the y direction, 
which shows the same peaks.

An interesting phenomenon happens when we calculate the spectrum
along the x direction, i.e., with $q_{||}\neq 0$ and $q_{\perp}=0$.
As predicted by 
eqs. (3), (4), we see one single
peak for each correlation function. 
Indeed,  as shown in figure 4, each 
power spectrum shows only one peak, and again as predicted by (3), (4),
the peak for the $v_{\perp}$
power spectrum is at a different $\omega$ than the peak of the density 
power spectrum!
This means that the velocity fluctuations propagate with a different 
velocity 
than the density fluctuations in the $x$-direction! 


We can then extract from
figure 4 the values of $v_s=0.93$, $\lambda=0.75$.
(The fact that $\lambda\ne 1$ reflects the absence of Galilean 
invariance). With the value of $c=0.62$ determined through figure 3, 
we can predict the sound speeds in all other
directions of propagation from eq. (6) with no adjustable
parameters. To test these predictions,
we have also calculated the power spectra
for the density and the velocity fields at two other angles:
$\tan (\theta_q)=1/3, 4$.
For the large angle $\theta_{q,1} =\arctan (4)=76.0^{o}$, the data are shown 
in fig. 5(a). 
The
peaks for $\rho$ and $v_{\perp}$ are at the same location, and the
wave velocities are $c_{\pm}(\theta_{q,1})=0.75,-0.37$. The data 
for $\theta_{q,2}=\arctan(1/3)=18.4^{o}$ are shown in fig. 5(b). The peak at
$\omega=c_{-}(\theta_{q,2})$ is just barely
visible in the density correlation, but both peaks show 
very well in the velocity correlation, and the peaks for both correlation 
functions are at the same locations, giving the velocity 
$c_{\pm}(\theta_{q,2})=0.97,0.59$. In fig. 5(c), we have plotted 
the angle dependence 
of the wave velocity as predicted in eq. (6) in polar angle
coordinates $(c_{\pm}(\theta_q),\theta_q)$, with the values of $v_s$, 
$\lambda$, and $c$ determined earlier. We have
included in fig. 5(c), the sound velocities for the two angles $\theta_{q,1}$
and $\theta_{q,2}$. The agreement with the predicted velocities is 
excellent.

In summary, 
the numerical simulations reported here strongly support our analytical
continuum theory of flocks. The observed sound speeds agree very well
with our predictions. In particular, our analytical model's
assertion that Galilean invariance is absent is confirmed
by the existence of two different non-zero sound speeds for
propagation along the mean direction of flock motion. 
In addition, the sound attenuation shows the 
anomalous scaling we predict\cite{foot2}.
 
Y. Tu is grateful to Dr. R. Walkup for helping with the parallel programming for 
this problem on an IBM SP-1 parallel computer.
J. Toner thanks the Aspen Center for Physics and the Center for Chaos and Complexity at the University of Colorado, Boulder for their hospitality while a portion of this work was completed.

\begin{figure}
\caption{(a)The order parameter $\phi$, as defined in the text, versus the
noise strength $\Delta v$. The arrow shows the value of $\Delta v$ at which the fluctuations
of the ordered state were calculated. (b)The scaling behavior of the
equal time correlation function for the density fluctuations in the two
limits as given in eqs. (9,10). The lines illustrate the predicted 
slopes. }
\label{fig1}
\end{figure}
\begin{figure}
\caption{The log-log plot of the anomalous transverse diffusion of an 
individual boid
versus time.}
\label{fig2}
\end{figure}
\begin{figure}
\caption{The power spectrum of the density for different
wave vectors. The inset shows the peak positions of the power spectrum 
versus wavenumber. The linear slope determines
the sound velocity.}
\label{fig3}
\end{figure}
\begin{figure}
\caption{Power spectra for the density and velocity fluctuations for the
same wave vector along the parallel direction. The peaks of the
two curves are clearly different.}

\label{fig4}
\end{figure}
\begin{figure}
\caption{The power spectra for the density and the velocity fluctuations
in directions (a)$\theta_{q,1}=\arctan(4)$ and (b) $\theta_{q,2}=\arctan(1/3)$.
The two peaks are clearly visible, albeit with different magnitudes. 
In (c), the
wave velocities $c_{\pm}(\theta_q)$ are plotted in
polar angle coordinates $(c_{\pm}(\theta_q),\theta_q)$ for the four different 
directions
$\theta_q=0, \theta_{q,1}, \theta_{q,2}, \pi/2$, the two axes represent 
$c_x=c_{\pm}(\theta_q)\cos (\theta_q)$ and $c_y=c_{\pm}(\theta_q)\sin (\theta_q)$
respectively. The solid curve is the prediction
from eq. (6) in the text.}
\label{fig5}

\end{figure}


\begin{references}
\bibitem{foot1}Additional non-linear terms $\lambda_2\vec{v}(\nabla\cdot
\vec{v})$ and $\lambda_3\nabla (|\vec{v}|^2)$ are also allowed, but
do not change any results in the broken symmetry state in $d=2$.
\bibitem{foot2}Our original derivation of the scaling exponents $z$, 
$\zeta$ and $\chi$ in ref. 3 assumed $\lambda=1$, which it need not be in 
general. However, we have shown in ref. 4 that the exponents are unchanged for a finite range of $\lambda$'s.
\end{references}
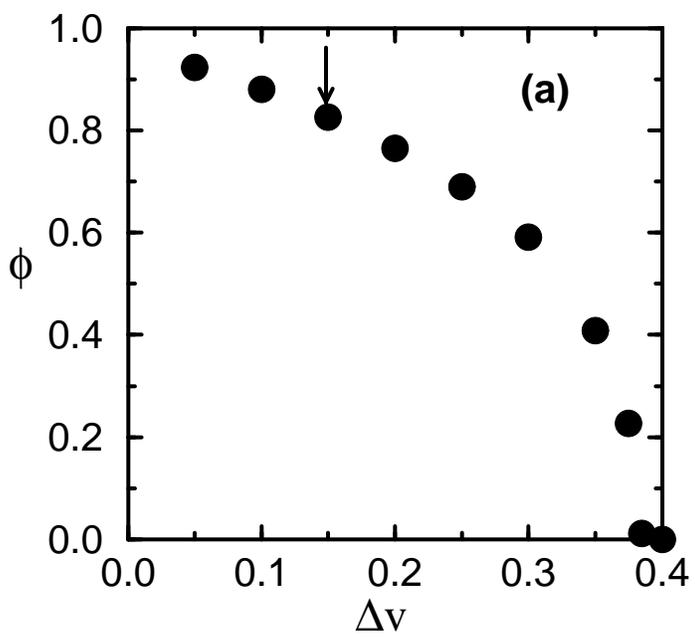
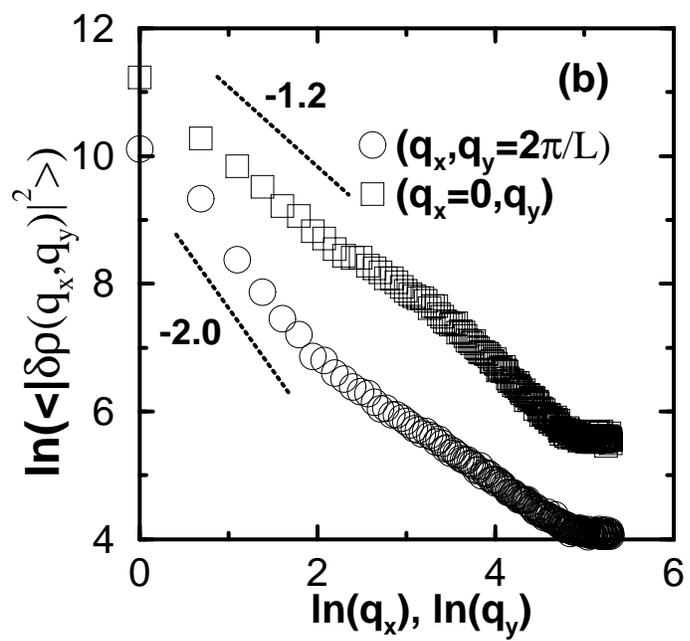
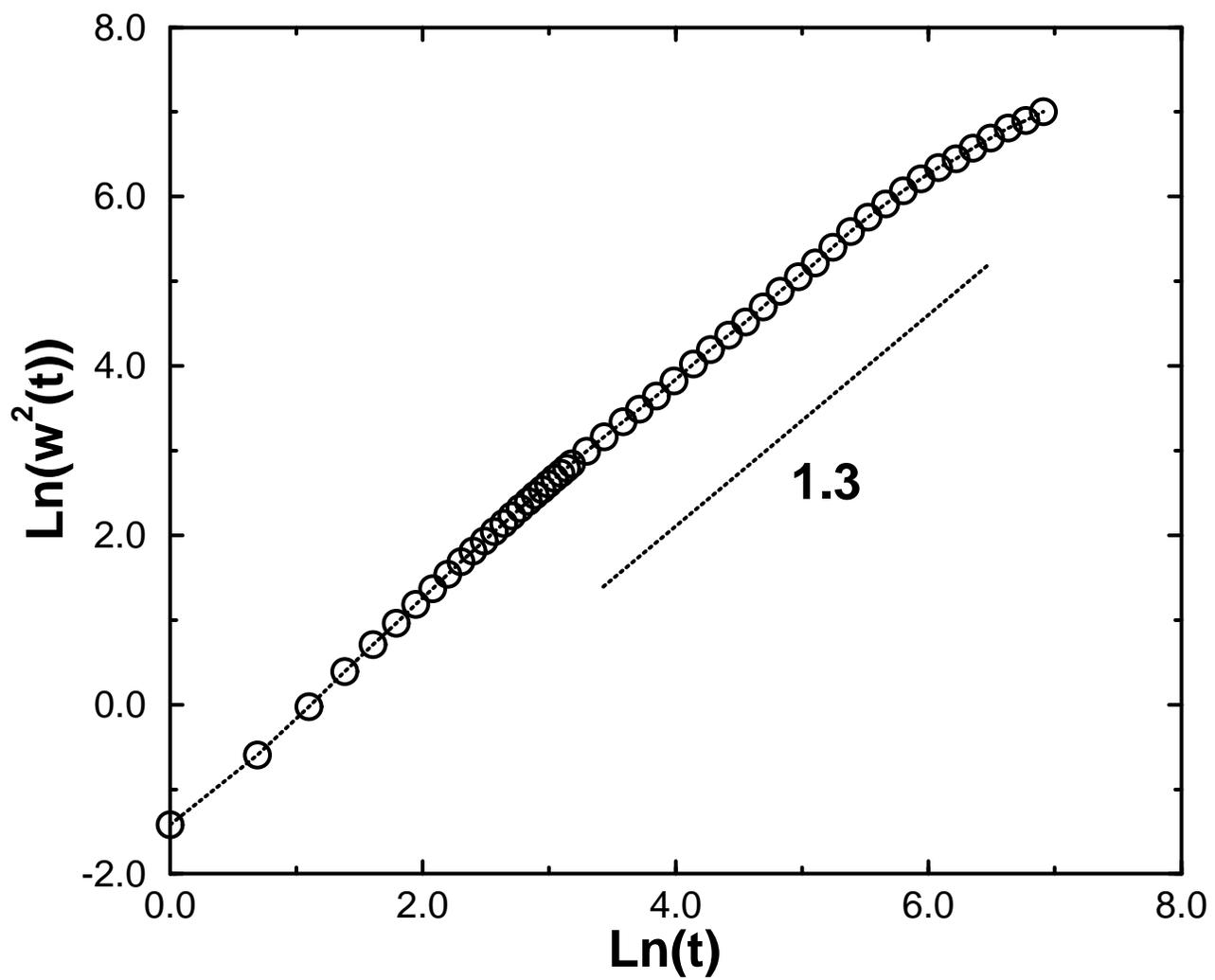
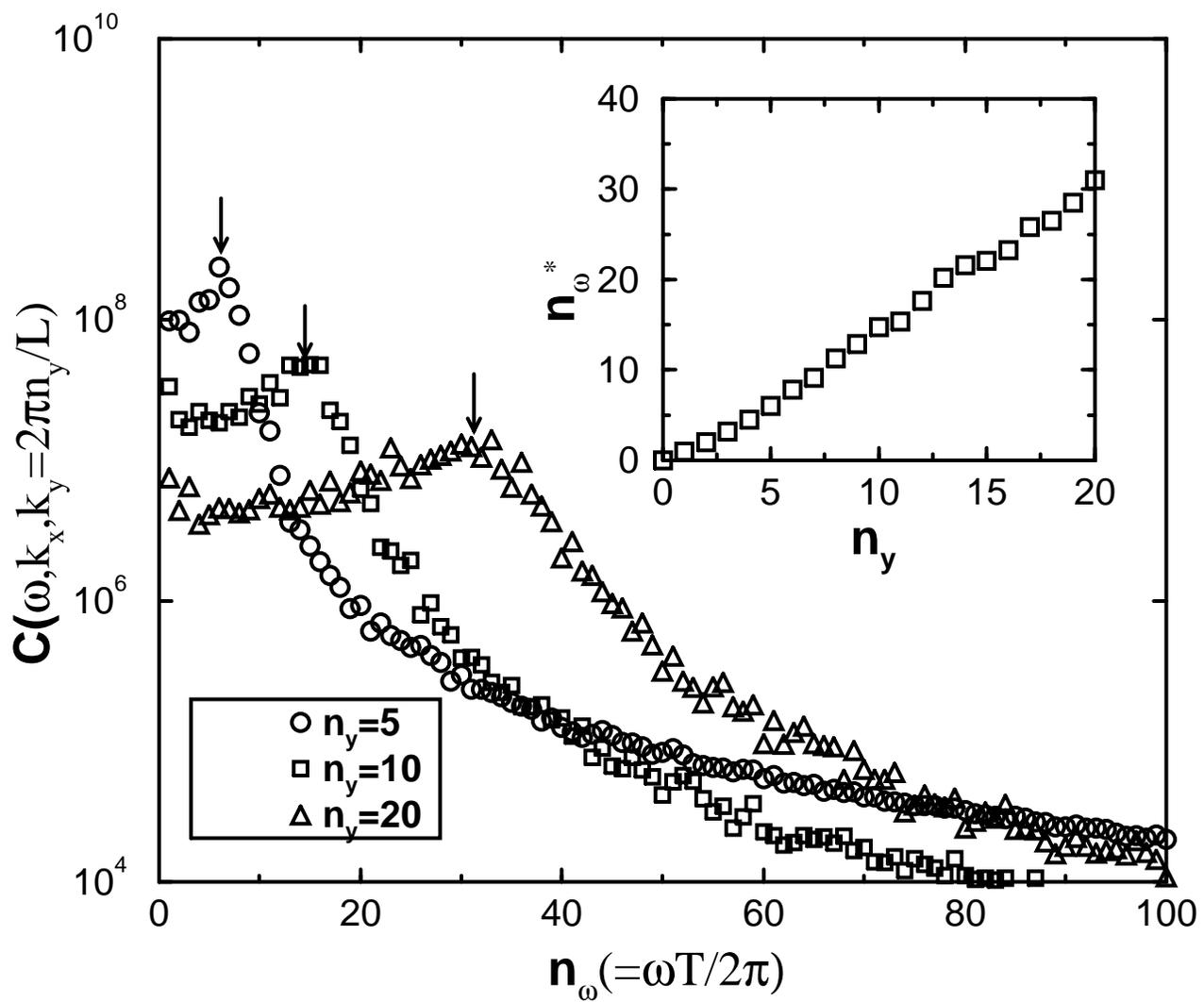
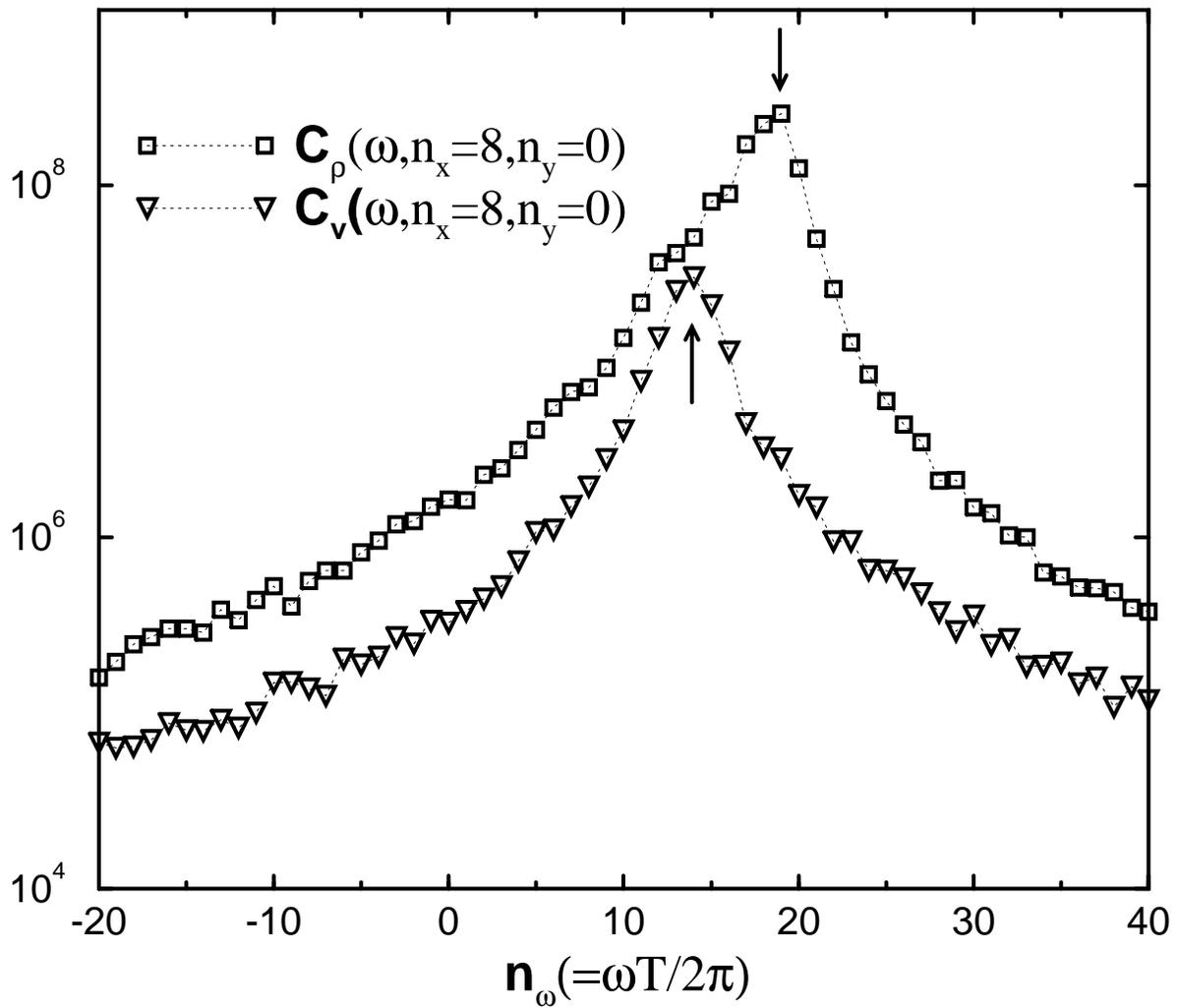
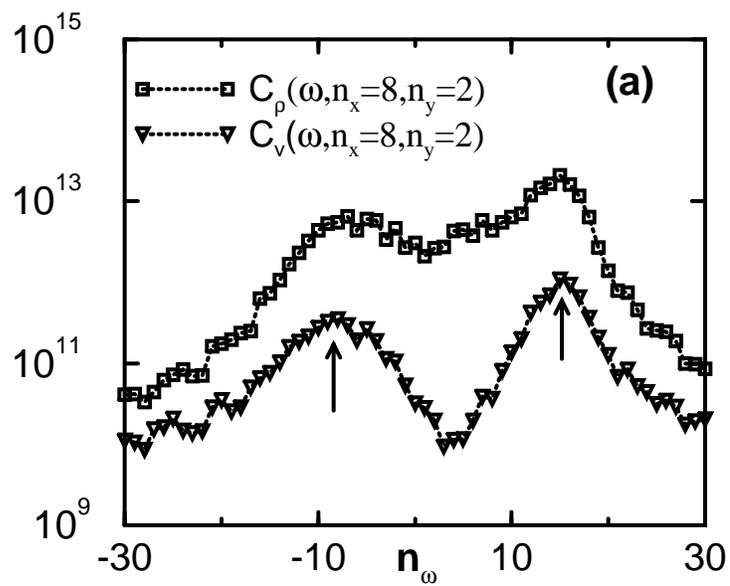
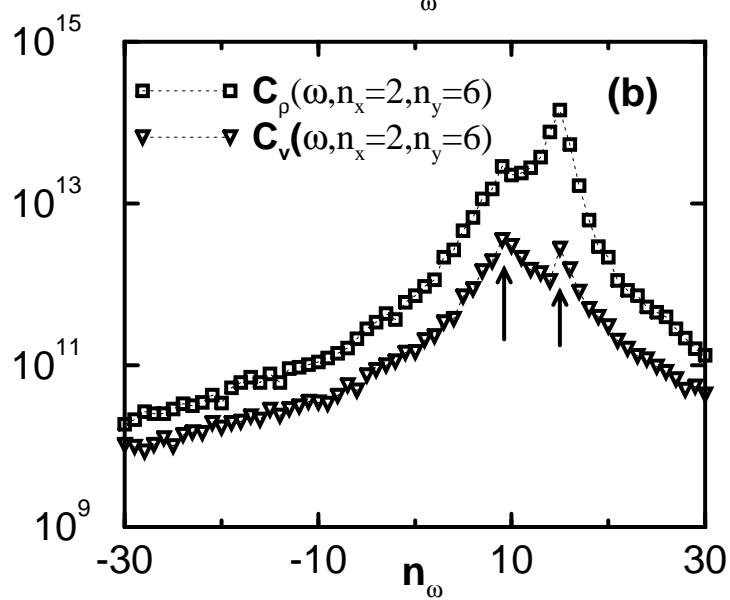
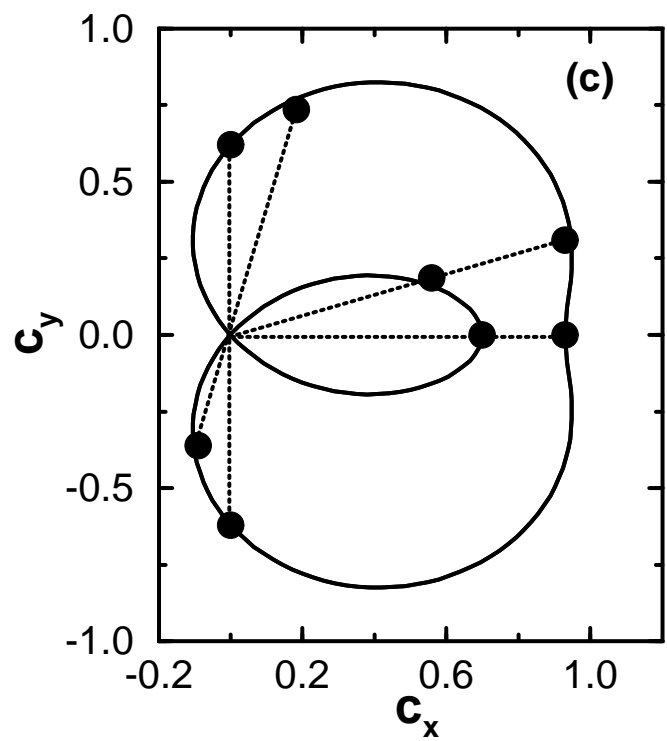
\end{document}